# CHIRALLY INVARIANT TRANSPORT EQUATIONS FOR QUARK MATTER[1]

W. Florkowski [2,3], J. Hüfner, S.P. Klevansky and L. Neise

Institut für Theoretische Physik, Universität Heidelberg
Philosophenweg 19, D-69120 Darmstadt, Germany

**Abstract:** Transport equations for quark matter are derived in the mean field approximation for the Nambu – Jona-Lasinio Lagrangian. Emphasis is put on the chiral limit (zero current quark mass) and the consequences of the chiral symmetry are investigated in detail. Our approach is based on the spinor decomposition of the Wigner function. Kinetic equations for the quark densities and for the evolution of the spin are derived. We find that the latter is constrained by the requirement of the axial current conservation.

HEIDELBERG, MAY 1995

[1] Research supported in part by the Federal Ministry for Research and Technology, (BMFT) grant number 06 HD 742 (0), the Deutsche Forschungsgemeinschaft, grant number Hu 233/4-3, and the Gesellschaft für Schwerionenforschung (GSI) m.b.H. Darmstadt.

[2] Permanent address: H. Niewodniczański Institute of Nuclear Physics, ul. Radzikowskiego 152, PL-31-342 Kraków, Poland.

[3] E-mail address: wojtek @ hobbit.mpi-hd.mpg.de.

## 1. Introduction

The ultimate goal of the ultra-relativistic heavy-ion collisions being currently carried out at CERN is the observation of the phase transition from hadronic matter to the so-called quark-gluon plasma [1, 2]. One expects that during such a deconfinement phase transition, in which the structures of individual hadrons are destroyed and replaced by a highly-energetic system of quarks and gluons, chiral symmetry is additionally restored. Lattice simulations of QCD [3] indicate that the chiral symmetry restoration phase transition and deconfinement phase transitions may indeed occur at the same temperature. Looking for the signatures of these two phase transitions is one of the most challenging problems of high-energy nuclear physics, both from the experimental and the theoretical point of view [4].

Since the matter produced in high energy nuclear collisions lives only for a short while, it is natural to expect that its space-time evolution proceeds far away from equilibrium. Consequently, there exists a growing interest in applying and developing transport theories, which are suitable for the description of non-equilibrium processes. A kinetic theory based directly on QCD has been formulated in papers by Heinz [5], and by Elze, Gyulassy and Vasak [6]. However this approach, with a few exceptions [7], is still far removed from practical applications.

In the context of the two phase transitions that have been mentioned, it is desirable and important to be able to include some of their aspects into the kinetic approach, which can otherwise be based on some phenomenological assumptions. The deconfinement phase transition seems at present to be too difficult to deal with, since we still do not yet understand this phenomenon fully. On the other hand, the chiral phase transition has been studied in terms of effective chiral models of QCD, for which phenomenological Lagrangians can be given. Such an approach has been quite successful in clarifying this transition.

One such effective approach is based on the Nambu – Jona-Lasinio (NJL) model; within its framework we can study the temperature (density) dependence of various physical quantitites like, e.g., the quark condensate, the pion decay constant, or the quark and meson masses [8, 9]. It is also possible to study the change of the thermodynamic properties of the quark-meson plasma as the system passes the chiral phase transition [10]. With few exceptions however [11, 12, 13], the considerations based on the NJL model are usually restricted to equilibrium situations. We refer the reader to reviews on the NJL model given in [14].

In the present paper, we wish to investigate how the concept of chiral symmetry can be *explicitly* included into the quantum and classical transport theory based on the NJL Lagrangian. In our investigations, we shall restrict ourselves to the mean field or Hartree approximation, since the interesting phenomena (like, e.g., spontanous symmetry breaking) already take place at this level. Our approach is based on the spinor decomposition of the Wigner function and follows the treatment of [15] and [16]. However, in contrast to [15] we do not neglect the spin degrees of freedom and take into account all the components of the spinor decomposition. On the other hand, we differ from [16], where the spin dynamics is fully discussed, in that we study a system that is governed by a different type of interaction.

A transport theory for the NJL model has been already formulated by Zhang and Wilets [11], where the Keldysh closed-time-path formalism combined with the effective-action method has been used to derive the kinetic equations. A certain class of solutions of the classical transport equations found in [11] has been studied in both [12] and [13]. Nevertheless, Ref. [11] does not discuss how the symmetry of the underlying theory (i.e., the chiral symmetry of the NJL Lagrangian) should reflect itself in the transport equations.

In this paper, in trying to clarify this point, we derive the general form of the transport equations (in the mean field approximation) which are chirally invariant. We discuss the properties of these equations and then study their classical limit by performing an expansion in $\hbar$ for all the functions appearing in the spinor decomposition. Throughout our approach, we investigate and check the consistency of our equations under the assumptions of chiral symmetry. In particular, this implies that the axial current should always be conserved. For the spin evolution equation, as will be seen, this turns out to be a non-trivial requirement.

Our paper is organized as follows: In Section 2 we define the NJL model in the mean field approximation. The Wigner function and its spinor decomposition are introduced in Section 3. In Section 4 we present the chirally invariant quantum kinetic equations. Their classical limit is discussed thoroughly in Sections 5 – 8. In Section 9 we study the case when chiral symmetry is explicitly broken by the presence of nonzero current quark masses. We summarize and conclude in Section 10.

## 2. The NJL model in the mean field approximation

In our study, we consider the following Lagrangian of the NJL type,

$$\mathcal{L} = \bar{\psi}\left(i\,\partial\!\!\!/ - m_0\right)\psi + G\left[(\bar{\psi}\psi)^2 + (\bar{\psi}i\gamma_5\psi)^2\right], \tag{1}$$

where $\psi$ is the Dirac field, $m_0$ is the current quark mass, and $G$ is the coupling constant. For simplicity, flavor and color degrees of freedom have been neglected. In the case $m_0 = 0$, the Lagrangian (1) is invariant under $U_A(1)$ transformations. This symmetry is, however, spontanously broken by the ground state. In a more realistic approach to quark matter, one should consider the Lagrange density which is invariant under $SU_A(2)$ transformations, with $U_A(1)$ explicitly broken due to the presence of anomalies and with the spontanous symmetry breaking of $SU_A(2)$ [14]. Nevertheless, we expect that the basic problems concerning the formulation of the chirally invariant transport theory can be already studied in the framework of a simpler theory, like this based on (1). *In what follows we shall always assume that $m_0 = 0$ except for Section 9 where the consequences of setting $m_0 \neq 0$ are taken into consideration.*

By introducing the fields $\hat{\sigma}$ and $\hat{\pi}$ defined as

$$\hat{\sigma} = -2G\,\bar{\psi}\psi, \quad \hat{\pi} = -2G\,\bar{\psi}i\gamma_5\psi, \qquad (2)$$

we can recast Lagrangian (1) into the form

$$\mathcal{L} = \bar{\psi}i\,\slashed{\partial}\psi - \hat{\sigma}\,\bar{\psi}\psi - \hat{\pi}\,\bar{\psi}i\gamma_5\psi - \frac{\hat{\sigma}^2 + \hat{\pi}^2}{4G}. \qquad (3)$$

The equivalence of (1) and (3) becomes clear if one checks that the variation of (1) with respect to $\bar{\psi}$ gives the same equation of motion for $\psi$ as the variation of (3) with respect to $\bar{\psi}, \hat{\sigma}$ and $\hat{\pi}$, namely

$$[i\,\slashed{\partial} - \hat{\sigma}(x) - i\gamma_5\hat{\pi}(x)]\,\psi(x) = 0. \qquad (4)$$

In the following, we shall restrict ourselves to the mean field approximation, i.e., the operators $\hat{\sigma}$ and $\hat{\pi}$ will be replaced by their mean values

$$\hat{\sigma} \to \sigma = \langle\hat{\sigma}\rangle = \text{Tr}(\hat{\rho}\hat{\sigma}), \quad \hat{\pi} \to \pi = \langle\hat{\pi}\rangle = \text{Tr}(\hat{\rho}\hat{\pi}). \qquad (5)$$

where $\hat{\rho}$ is the density operator and Tr denotes the trace over all physical states of the system. The particular form of $\hat{\rho}$ is not important for our considerations. We only note that in typical situations, $\hat{\rho}$ is either an operator projecting on the parity invariant ground state or it describes parity invariant statistical ensembles. In these two cases $\langle\hat{\pi}\rangle = 0$. Nevertheless, in more general situations (e.g., when the system is out of equilibrium) we cannot exclude the case $\langle\hat{\pi}\rangle \neq 0$.

The Lagrangian (3) is invariant under the $U_V(1)$ transformations, $\psi \to \psi' = \exp(-i\frac{\phi}{2})\psi$, which leads to the conservation of the vector current,

$$\partial_\mu V^\mu(x) = 0, \quad V^\mu(x) = \langle\bar{\psi}(x)\gamma^\mu\psi(x)\rangle. \qquad (6)$$

In addition, as was stated before, it is invariant under $U_A(1)$ chiral transformations

$$\psi \;\to\; \psi' = \exp(-i\gamma_5 \frac{\chi}{2})\psi, \tag{7}$$

$$\hat{\sigma} \;\to\; \hat{\sigma}' = \hat{\sigma}\cos\chi - \hat{\pi}\sin\chi, \tag{8}$$

$$\hat{\pi} \;\to\; \hat{\pi}' = \hat{\pi}\cos\chi + \hat{\sigma}\sin\chi. \tag{9}$$

Consequently, the axial current is conserved as well,

$$\partial_\mu A^\mu(x) = 0, \quad A^\mu(x) = \langle \bar{\psi}(x)\gamma^\mu\gamma_5\psi(x)\rangle. \tag{10}$$

In general, for a non-equilibrium transport theory, one requires two independent Green functions to describe the system. In the mean field approximation, however, one is sufficient. We thus choose the Green function

$$G^<_{\alpha\beta}(x,y) = \langle \bar{\psi}_\beta(y)\psi_\alpha(x)\rangle \tag{11}$$

to be the fundamental quantity in our description. One can easily notice that it satisfies the same equation as the field $\psi(x)$ does, i.e., in the mean field approximation we have

$$[i\,\slashed{\partial} - \sigma(x) - i\gamma_5\pi(x)]\, G^<(x,y) = 0. \tag{12}$$

The mean fields $\sigma(x)$ and $\pi(x)$ are determined by the Green function $G^<(x,y)$ through the relations

$$\sigma(x) = -2G\,\mathrm{tr}\, G^<(x,x), \quad \pi(x) = -2G\,\mathrm{tr}\, i\gamma_5 G^<(x,x), \tag{13}$$

where the trace tr is taken over the spinor indices.

### 3. Wigner function and its spinor decomposition

For the description of non-uniform systems, it is convenient to introduce Wigner transforms. This allows one to make an easier physical interpretation of the components of the Green function and facilitates performing the classical limit. In the case of the Green function $G^<(x,y)$, its Wigner transform is obtained by introducing the center-of-mass coordinate $X = (x+y)/2$, the relative coordinate $u = x - y$, and by taking the Fourier transform with respect to $u$. Following the notation of [17], we define

$$W_{\alpha\beta}(X,p) = \int d^4u\, e^{\frac{i}{\hbar}p\cdot u}\, G^<_{\alpha\beta}\left(X + \frac{u}{2}, X - \frac{u}{2}\right). \tag{14}$$

In Eq. (14) we have explicitly incorporated the Planck constant $\hbar$, since later on we wish to investigate the classical approximation.

The Wigner transforms of the derivative of a two-point function, $\partial f(x,y)/\partial x^\mu$, and of the product of a one-point function with the two-point one, $f(x)g(x,y)$, are given by the expressions

$$\frac{\partial f(x,y)}{\partial x^\mu} \rightarrow (-ip^\mu + \frac{\hbar}{2}\partial^\mu)f(X,p), \tag{15}$$

$$f(x)g(x,y) \rightarrow f(X)g(X,p) - \frac{i\hbar}{2}\partial_\mu f(X)\partial_p^\mu g(X,p), \tag{16}$$

with the notation $\partial_\mu = \partial/\partial X^\mu, \partial_p^\mu = \partial/\partial p_\mu$, and where in the second relation (16), only the derivatives of first order in $\partial_p$ and $\partial_X$ are kept. This approximation will always be used in the following.

Using Eq. (12) and the properties of the Wigner transform (15) and (16), we arrive at the following equation for the Wigner function

$$\left[\left(p^\mu + \frac{i\hbar}{2}\partial^\mu\right)\gamma_\mu - \sigma(X) + \frac{i\hbar}{2}\partial_\mu\sigma(X)\partial_p^\mu - i\gamma_5\pi(X) - \frac{\hbar}{2}\gamma_5\partial_\mu\pi(X)\partial_p^\mu\right]W(X,p) = 0. \tag{17}$$

Eq. (17) together with Eqs. (13) and (14) form the system of the coupled equations that we are going to study in detail in this paper. We observe that under the chiral transformation (7), the Wigner function changes according to the prescription

$$W \rightarrow W' = \exp(-i\gamma_5\frac{\chi}{2})\ W\ \exp(-i\gamma_5\frac{\chi}{2}). \tag{18}$$

Hence, one can check that Eq. (17) is invariant under transformations (8), (9) and (18) and thus that the first order in the derivative expansion of the Green function is symmetry preserving.

The Wigner function $W_{\alpha\beta}(X,p)$ is a 4 by 4 matrix in spinor indices and can be expanded in terms of 16 independent generators of the Clifford algebra. The conventional basis consists of $1, i\gamma_5, \gamma^\mu, \gamma^\mu\gamma_5$, and $\frac{1}{2}\sigma^{\mu\nu}$ which we denote in what follows by $\Gamma_i$. In this basis, the Wigner function has the form

$$W = \mathcal{F} + i\gamma_5\mathcal{P} + \gamma^\mu\mathcal{V}_\mu + \gamma^\mu\gamma_5\mathcal{A}_\mu + \frac{1}{2}\sigma^{\mu\nu}\mathcal{S}_{\mu\nu}. \tag{19}$$

The properties $\gamma^0 W^\dagger \gamma^0 = W$ and $\gamma^0 \Gamma_i^\dagger \gamma^0 = \Gamma_i$ (the first one results from (14), whereas the second one is a simple consequence of the definition of $\Gamma_i$) indicate that the coefficients in the spinor decomposition $\mathcal{F}(X,p), \mathcal{P}(X,p), \mathcal{V}_\mu(X,p), \mathcal{A}_\mu(X,p)$ and $\mathcal{S}_{\mu\nu}(X,p)$ are real functions. We note that the spinor decomposition technique has been already used in the formulation of the transport theory for QED [16, 18, 19] and for QHD (so-called quantum hadrodynamics) [15].

The vector and axial currents, see Eqs. (6) and (10), are related to the functions $\mathcal{V}^\mu(X,p)$ and $\mathcal{A}^\mu(X,p)$ by the simple relations

$$V^\mu(X) = 4 \int \frac{d^4p}{(2\pi)^4} \mathcal{V}^\mu(X,p) \tag{20}$$

and

$$A^\mu(X) = -4 \int \frac{d^4p}{(2\pi)^4} \mathcal{A}^\mu(X,p). \tag{21}$$

Therefore $\mathcal{V}^\mu(X,p)$ and $\mathcal{A}^\mu(X,p)$ can be interpreted as the momentum densities of these two currents. For the physical interpretation of the other components we refer the reader to Refs. [16, 18]. Under the chiral transformation (18) the coefficients of the spinor decomposition (19) change as follows

$$\begin{aligned}
\mathcal{F} &\to \mathcal{F}' = \mathcal{F} \cos\chi + \mathcal{P} \sin\chi, \\
\mathcal{P} &\to \mathcal{P}' = -\mathcal{F} \sin\chi + \mathcal{P} \cos\chi, \\
\mathcal{V}_\mu &\to \mathcal{V}'_\mu = \mathcal{V}_\mu, \\
\mathcal{A}_\mu &\to \mathcal{A}'_\mu = \mathcal{A}_\mu, \\
\mathcal{S}_{\mu\nu} &\to \mathcal{S}'_{\mu\nu} = \mathcal{S}_{\mu\nu} \cos\chi + \tilde{\mathcal{S}}_{\mu\nu} \sin\chi, \\
\tilde{\mathcal{S}}_{\mu\nu} &\to \tilde{\mathcal{S}}'_{\mu\nu} = -\mathcal{S}_{\mu\nu} \sin\chi + \tilde{\mathcal{S}}_{\mu\nu} \cos\chi,
\end{aligned} \tag{22}$$

where $\tilde{\mathcal{S}}^{\mu\nu}$ is the dual tensor to $\mathcal{S}^{\mu\nu}$, namely

$$\tilde{\mathcal{S}}^{\mu\nu} = \frac{1}{2} \varepsilon^{\mu\nu\alpha\beta} \mathcal{S}_{\alpha\beta}. \tag{23}$$

## 4. Kinetic equations

Substituting expression (19) into Eq. (17) gives the following system of the coupled equations for the coefficients of the decomposition (19)

$$K^\mu \mathcal{V}_\mu - \sigma \mathcal{F} + \pi \mathcal{P} = -\frac{i\hbar}{2}\left(\partial_\nu \sigma \partial_p^\nu \mathcal{F} - \partial_\nu \pi \partial_p^\nu \mathcal{P}\right), \tag{24}$$

$$-iK^\mu \mathcal{A}_\mu - \sigma \mathcal{P} - \pi \mathcal{F} = -\frac{i\hbar}{2}\left(\partial_\nu \sigma \partial_p^\nu \mathcal{P} + \partial_\nu \pi \partial_p^\nu \mathcal{F}\right), \tag{25}$$

$$K_\mu \mathcal{F} + iK^\nu \mathcal{S}_{\nu\mu} - \sigma \mathcal{V}_\mu + i\pi \mathcal{A}_\mu = -\frac{i\hbar}{2}\left(\partial_\nu \sigma \partial_p^\nu \mathcal{V}_\mu - i\partial_\nu \pi \partial_p^\nu \mathcal{A}_\mu\right), \tag{26}$$

$$iK^\mu \mathcal{P} - K_\nu \tilde{\mathcal{S}}^{\nu\mu} - \sigma \mathcal{A}^\mu + i\pi \mathcal{V}^\mu = -\frac{i\hbar}{2}\left(\partial_\nu \sigma \partial_p^\nu \mathcal{A}^\mu - i\partial_\nu \pi \partial_p^\nu \mathcal{V}^\mu\right), \tag{27}$$

$$i(K^\mu \mathcal{V}^\nu - K^\nu \mathcal{V}^\mu) - \varepsilon^{\mu\nu\alpha\beta} K_\alpha \mathcal{A}_\beta - \pi \tilde{\mathcal{S}}^{\mu\nu} + \sigma \mathcal{S}^{\mu\nu} = \frac{i\hbar}{2}(\partial_\gamma \sigma \partial_p^\gamma \mathcal{S}^{\mu\nu} - \partial_\gamma \pi \partial_p^\gamma \tilde{\mathcal{S}}^{\mu\nu}). \tag{28}$$

In abbreviation we have introduced here the notation $K^\mu = p^\mu + \frac{i\hbar}{2}\partial^\mu$. Eqs. (24) - (28) should be supplemented by the formulas that determine the mean fields. They follow from (13) and have the form

$$\sigma(X) = -2G \int \frac{d^4p}{(2\pi)^4} \operatorname{tr} W(X,p) = -8G \int \frac{d^4p}{(2\pi)^4} \mathcal{F}(X,p) \tag{29}$$

and

$$\pi(X) = -2G \int \frac{d^4p}{(2\pi)^4} \operatorname{tr} i\gamma_5 W(X,p) = 8G \int \frac{d^4p}{(2\pi)^4} \mathcal{P}(X,p). \tag{30}$$

Using the transformation properties (8), (9) and (22) we can check that this system of equations is chirally invariant. Strictly speaking, each of Eqs. (24), (25) and (28) is separately a chirally invariant equation. On the other hand, Eqs. (26) and (27), after a chiral transformation, form linear combinations of each other. These combinations imply, however, that each of their components must vanish separately. In consequence, we can treat *the system of the two* equations (26) and (27) as a chirally invariant expression. In the analogous way, one can check that the system of Eqs. (29) and (30) is chirally invariant.

For further analysis, it is convenient to discuss separately the real and imaginary parts of Eqs. (24) - (28). The real parts give

$$p^\mu \mathcal{V}_\mu - \sigma \mathcal{F} + \pi \mathcal{P} = 0, \tag{31}$$

$$\frac{\hbar}{2}\partial_\mu \mathcal{A}^\mu - \sigma\mathcal{P} - \pi\mathcal{F} = 0, \tag{32}$$

$$p_\mu \mathcal{F} - \frac{\hbar}{2}\partial^\nu \mathcal{S}_{\nu\mu} - \sigma\mathcal{V}_\mu = -\frac{\hbar}{2}\partial_\nu \pi \partial_p^\nu \mathcal{A}_\mu, \tag{33}$$

$$-\frac{\hbar}{2}\partial^\mu \mathcal{P} - p_\nu \tilde{\mathcal{S}}^{\nu\mu} - \sigma\mathcal{A}^\mu = -\frac{\hbar}{2}\partial_\nu \pi \partial_p^\nu \mathcal{V}^\mu, \tag{34}$$

$$-\frac{\hbar}{2}\left(\partial^\mu \mathcal{V}^\nu - \partial^\nu \mathcal{V}^\mu\right) - \varepsilon^{\mu\nu\alpha\beta} p_\alpha \mathcal{A}_\beta + \sigma\mathcal{S}^{\mu\nu} - \pi\tilde{\mathcal{S}}^{\mu\nu} = 0. \tag{35}$$

One can notice that integrating of Eq. (32) over the momentum gives

$$\hbar \int \frac{d^4p}{(2\pi)^4} \partial_\mu \mathcal{A}^\mu(X,p) = 0, \tag{36}$$

where the definitions (29) and (30) have been used. This equation is nothing other than the statement that the axial current should be conserved (10). We thus see that after making the gradient expansion, this conservation law is still included in the transport equations. The imaginary parts of Eqs. (24) - (28) yield

$$\frac{\hbar}{2}\partial^\mu \mathcal{V}_\mu = -\frac{\hbar}{2}\left(\partial_\nu \sigma \partial_p^\nu \mathcal{F} - \partial_\nu \pi \partial_p^\nu \mathcal{P}\right), \tag{37}$$

$$p^\mu \mathcal{A}_\mu = \frac{\hbar}{2}\left(\partial_\nu \sigma \partial_p^\nu \mathcal{P} + \partial_\nu \pi \partial_p^\nu \mathcal{F}\right), \tag{38}$$

$$\frac{\hbar}{2}\partial_\mu \mathcal{F} + p^\nu \mathcal{S}_{\nu\mu} + \pi\mathcal{A}_\mu = -\frac{\hbar}{2}\partial_\nu \sigma \partial_p^\nu \mathcal{V}_\mu, \tag{39}$$

$$p^\mu \mathcal{P} - \frac{\hbar}{2}\partial_\nu \tilde{\mathcal{S}}^{\nu\mu} + \pi\mathcal{V}^\mu = -\frac{\hbar}{2}\partial_\nu \sigma \partial_p^\nu \mathcal{A}^\mu, \tag{40}$$

$$(p^\mu \mathcal{V}^\nu - p^\nu \mathcal{V}^\mu) - \frac{\hbar}{2}\varepsilon^{\mu\nu\alpha\beta}\partial_\alpha \mathcal{A}_\beta = \frac{\hbar}{2}\left(\partial_\gamma \sigma \partial_p^\gamma \mathcal{S}^{\mu\nu} - \partial_\gamma \pi \partial_p^\gamma \tilde{\mathcal{S}}^{\mu\nu}\right). \tag{41}$$

Although we have neglected the higher order gradients, Eqs. (31) - (41) are still quantum kinetic equations. In order to obtain the classical equations one makes an expansion in powers of $\mathcal{F}$ in $\hbar$

$$\mathcal{F} = \mathcal{F}_{(0)} + \hbar \mathcal{F}_{(1)} + \hbar^2 \mathcal{F}_{(2)} + ..., \tag{42}$$

and similarly of $\mathcal{P}, \mathcal{V}^\mu, \mathcal{A}^\mu, \mathcal{S}^{\mu\nu}, \pi$ and $\sigma$. Expansions of the form (42) are inserted into Eqs. (31) - (41) and the expressions at the appropriate powers of $\hbar$ are compared. The detailed description of this procedure will be the subject of the next Sections.

## 5. Constraint equations in the leading order of $\hbar$

Substituting expansions of the type (42) into Eqs. (31) - (35), we find to leading (zeroth) order of $\hbar$

$$p^\mu \mathcal{V}_\mu^{(0)} - \sigma_{(0)} \mathcal{F}_{(0)} + \pi_{(0)} \mathcal{P}_{(0)} = 0, \tag{43}$$

$$\sigma_{(0)} \mathcal{P}_{(0)} + \pi_{(0)} \mathcal{F}_{(0)} = 0, \tag{44}$$

$$p^\mu \mathcal{F}_{(0)} - \sigma_{(0)} \mathcal{V}_{(0)}^\mu = 0, \tag{45}$$

$$\tilde{\mathcal{S}}_{(0)}^{\mu\nu} p_\nu - \sigma_{(0)} \mathcal{A}_{(0)}^\mu = 0, \tag{46}$$

$$-\varepsilon^{\mu\nu\alpha\beta} p_\alpha \mathcal{A}_\beta^{(0)} + \sigma_{(0)} \mathcal{S}_{(0)}^{\mu\nu} - \pi \tilde{\mathcal{S}}_{(0)}^{\mu\nu} = 0. \tag{47}$$

Correspondingly, the zeroth order of Eqs. (37) - (41) has the form

$$p^\mu \mathcal{A}_\mu^{(0)} = 0, \tag{48}$$

$$p^\nu \mathcal{S}_{\nu\mu}^{(0)} + \pi^{(0)} \mathcal{A}_\mu^{(0)} = 0, \tag{49}$$

$$p^\mu \mathcal{P}^{(0)} + \pi^{(0)} \mathcal{V}_{(0)}^\mu = 0, \tag{50}$$

$$p^\mu \mathcal{V}_{(0)}^\nu - p^\nu \mathcal{V}_{(0)}^\mu = 0. \tag{51}$$

These are only four equations, since Eq. (37) is already of first order in $\hbar$. From Eqs. (44) and (45) one finds

$$\mathcal{P}_{(0)} = -\pi_{(0)} \frac{\mathcal{F}_{(0)}}{\sigma_{(0)}} \tag{52}$$

and

$$\mathcal{V}_{(0)}^\mu = p^\mu \frac{\mathcal{F}_{(0)}}{\sigma_{(0)}}. \tag{53}$$

Then Eqs. (50) and (51) are automatically fulfilled. Moreover, Eqs. (43) - (45) lead to the mass-shell constraint for the function $\mathcal{F}_{(0)}(X,p)$, namely

$$[p^2 - M^2(X)] \mathcal{F}_{(0)}(X,p) = 0, \quad M^2(X) = \sigma_{(0)}^2(X) + \pi_{(0)}^2(X). \tag{54}$$

From Eq. (47), we find the following expression for the spin tensor

$$\mathcal{S}_{(0)}^{\mu\nu} = -\frac{\pi_{(0)}}{M^2} \left[ p^\mu \mathcal{A}_{(0)}^\nu - p^\nu \mathcal{A}_{(0)}^\mu \right] + \frac{\sigma_{(0)}}{M^2} \varepsilon^{\mu\nu\alpha\beta} p_\alpha \mathcal{A}_\beta^{(0)}, \tag{55}$$

and the dual spin tensor

$$\tilde{\mathcal{S}}_{(0)}^{\mu\nu} = -\frac{\sigma_{(0)}}{M^2}\left[p^\mu \mathcal{A}_{(0)}^\nu - p^\nu \mathcal{A}_{(0)}^\mu\right] - \frac{\pi_{(0)}}{M^2}\varepsilon^{\mu\nu\alpha\beta}p_\alpha \mathcal{A}_\beta^{(0)}. \tag{56}$$

Substituting now Eq. (56) into Eq. (46), and using (48) one obtains a mass-shell constraint for $\mathcal{A}_{(0)}^\mu(X,p)$ also, i.e.,

$$[p^2 - M^2(X)]\mathcal{A}_{(0)}^\nu(X,p) = 0. \tag{57}$$

Finally, using (55), (48) and (57) we find that (49) is satisfied. In summary, Eqs. (43) - (51) lead to the expressions for $\mathcal{P}_{(0)}$ and $\mathcal{V}_{(0)}^\mu$ in terms of $\mathcal{F}_{(0)}$, and for $\mathcal{S}_{(0)}^{\mu\nu}$ and $\tilde{\mathcal{S}}_{(0)}^{\mu\nu}$ in terms of $\mathcal{A}_{(0)}^\mu$. Furthermore, they lead to the two mass-shell constraints.

The mean fields appearing in Eqs. (43) - (47), (49) and (50) are required to be calculated in a self-consistent way from Eqs. (29) and (30). Doing so, we find

$$\sigma_{(0)}(X) = -8G \int \frac{d^4p}{(2\pi)^4} \mathcal{F}_{(0)}(X,p), \tag{58}$$

$$\pi_{(0)}(X) = 8G \int \frac{d^4p}{(2\pi)^4} \mathcal{P}_{(0)}(X,p). \tag{59}$$

Due to the relation (44), one finds that these two equations are not independent and can be reduced to a single equation which determines the *invariant* mass $M(X)$. To see this relation more clearly, let us define the distribution function $F(X,p)$ via the expression

$$F(X,p) = \frac{\mathcal{F}_{(0)}(X,p)}{\sigma_{(0)}(X)} = -\frac{\mathcal{P}_{(0)}(X,p)}{\pi_{(0)}(X)} \tag{60}$$

and the angle $\Phi(X)$ via the relations

$$\pi_{(0)} = M(X)\sin\Phi(X), \quad \sigma_{(0)} = M(X)\cos\Phi(X). \tag{61}$$

$F(X,p)$ is a chirally invariant quantity, and knowledge of it determines the value of $M(X)$. In addition, it allows us to calculate the vector current density $\mathcal{V}_{(0)}^\mu$. However, Eqs. (58) and (59) do not determine separately the fields $\pi_{(0)}$ and $\sigma_{(0)}$, i.e., the angle $\Phi(X)$. This fact is in agreement with the requirements of the chiral symmetry of the problem: under the chiral transformations $\Phi(X) \to \Phi'(X) = \Phi(X) + \chi$ and, consequently, the absolute value of $\Phi(X)$ has no physical significance.

With this, we conclude the discussion of the expressions obtained to zeroth order in $\hbar$. The derived equations all display the chirally invariant form. Let us turn now to the discussion of the equations which follow from Eqs. (37) - (41), considered to first order in $\hbar$.

## 6. Kinetic equation for the quark distribution functions

Employing Eqs. (52) and (53) in Eq. (37), one immediately finds

$$p^\mu \partial_\mu F(X,p) + M(X) \partial_\mu M(X) \partial_p^\mu F(X,p) = 0, \tag{62}$$

where we have used the definition (60). This is again a chirally invariant equation. Due to the mass-shell condition (54), we can express $F(X,p)$ as the sum of the quark and antiquark distribution functions $f^+(X,\mathbf{p})$ and $f^-(X,\mathbf{p})$ defined by

$$F(X,p) = 2\pi \left\{ \frac{\delta(p^0 - E_p(X))}{2E_p(X)} f^+(X,\mathbf{p}) + \frac{\delta(p^0 + E_p(X))}{2E_p(X)} \left[ f^-(X,-\mathbf{p}) - 1 \right] \right\}. \tag{63}$$

Here $E_p(X) = \sqrt{\mathbf{p}^2 + M^2(X)}$, and $\mathbf{p}$ is a three-momentum. Substituting expression (63) into Eq. (62), and integrating over $p^0$ gives

$$p^\mu \partial_\mu f^\pm(X,\mathbf{p}) + M(X) \partial_\mu M(X) \partial_p^\mu f^\pm(X,\mathbf{p}) = 0. \tag{64}$$

Eq. (64) displays a typical form for fermion mean field theories. In our case, however, a characteristic feature of this equation is the appearance of the chirally invariant mass $M(X)$. We emphasize that the four-momentum $p^\mu$ occuring in (64) is to be taken on the mass shell, i.e., in this case $p^0 = E_p(X)$. At the same time the distribution functions $f^\pm$ are functions of the space-time coordinate $X$ and the three-momentum $\mathbf{p}$ with the derivative with respect to the energy, $p^0$, defined to be zero. We have formally included it however, in order to display the relativistic invariance of the kinetic equation (64) explicitly.

Using Eqs. (58), (60) and (63) we find that the NJL gap equation takes the form

$$\sigma_{(0)}(X) = 4G\sigma_{(0)}(X) \int \frac{d^3p}{(2\pi)^3} \frac{1}{E_p(X)} \left[ 1 - f^+(X,\mathbf{p}) - f^-(X,\mathbf{p}) \right], \tag{65}$$

which determines the mass $M(X)$ appearing in $E_p(X)$ in terms of the distribution functions $f^\pm(X,\mathbf{p})$. One can easily notice that using Eq. (59) instead of (58) we would obtain the same equation as (65), with the appearance of $\pi_{(0)}$ instead of $\sigma_{(0)}$.

Using now Eqs. (20) and (63), we find the following expression for the baryon current

$$V^\mu(X) = 2 \int \frac{d^3p}{(2\pi)^3} \frac{p^\mu}{E_p(X)} \left[ f^+(X,\mathbf{p}) - f^-(X,\mathbf{p}) + 1 \right], \tag{66}$$

where $p^\mu$ is again on the mass shell, i.e., $p^0 = E_p(X)$. The constant appearing at the end of the square bracket can be neglected, since it does not contribute to $\mathbf{V}(X)$, and for $V^0(X)$ it introduces a constant (although infinite) charge.

We note that Eqs. (64) and (65) form a closed system of equations: the distribution functions $f^\pm(X,p)$ determine the invariant mass $M(X)$ through Eq. (65), whereas the space dependence of $M(X)$ determines in (64) the time evolution of $f^\pm(X,p)$. In the case $\pi_{(0)} = 0$ this system of equations was first derived by Zhang and Wilets [11], whose calculations were based on the path integral formalism. Our approach represents an alternative derivation and generalizes their results to the case $\pi_{(0)} \neq 0$. Some solutions of Eqs. (64) and (65) have been found in [12] and [13].

## 7. Spin evolution

In the classical limit the spin dynamics is described by the behavior of the function $\mathcal{A}^\mu_{(0)}(X,p)$ [16, 18]. In order to derive the kinetic equation for $\mathcal{A}^\mu_{(0)}(X,p)$, we use Eq. (40) and substitute into it the expression for $\mathcal{V}^\mu$ obtained from Eq. (33). Such a procedure gives, to the first order in $\hbar$, the following equation

$$p^\mu \partial_\nu \mathcal{A}^\nu_{(0)} - \sigma_{(0)} \partial_\nu \tilde{\mathcal{S}}^{\nu\mu}_{(0)} - \pi_{(0)} \partial_\nu \mathcal{S}^{\nu\mu}_{(0)} + M \partial_\nu M \partial_p^\nu \mathcal{A}^\mu_{(0)} = 0. \tag{67}$$

Substituting the expression for the spin tensor (55) and the dual spin tensor (56) into Eq. (67) leads to the equation determining the space-time evolution of $\mathcal{A}^\mu_{(0)}(X,p)$, namely

$$p^\nu \partial_\nu \mathcal{A}^\mu_{(0)} + M \partial_\nu M \partial_p^\nu \mathcal{A}^\mu_{(0)} + \frac{\partial_\nu M}{M} \left[ p^\mu \mathcal{A}^\nu_{(0)} - p^\nu \mathcal{A}^\mu_{(0)} \right] - \varepsilon^{\mu\nu\alpha\beta} \partial_\nu \Phi p_\alpha \mathcal{A}^{(0)}_\beta = 0. \tag{68}$$

We note that Eq. (68) is again a chirally invariant equation, since both the mass $M(X)$ and the gradient of the angle $\Phi(X)$ are chirally invariant quantities. Due to the condition (48), only three out of four equations in (68) are independent. In fact, one can check that multiplication of (68) by $p^\mu$ gives zero.

The mass $M(X)$ appearing in (68) has to be calculated from the system of equations (64) and (65). Hence, it can be treated in (68) as an externally prescribed function. On the other hand, the gradient $\partial_\mu \Phi(X)$ is not known, and until we specify how to calculate it, Eq. (68) cannot be used to determine the time evolution of $\mathcal{A}^\mu_{(0)}(X,p)$. Moreover, one has to check whether the solutions of (68) satisfy the requirement of axial current conservation (10). The last two points will be discussed in more detail in the next Section, where we examine the consistency of our all equations up to the first order in $\hbar$.

In analogy to the QED calculations [6] we introduce the spin up and spin down phase-space densities

$$F_{\pm s}(X,p) = F(X,p) \pm S_\mu(X,p) \frac{\mathcal{A}^\mu_{(0)}(X,p)}{M(X)}, \tag{69}$$

where $S_\mu(X,p)$ is defined by

$$S_\mu(X,p) = \frac{\mathcal{A}^{(0)}_\mu(X,p)}{\left[-\mathcal{A}^{(0)}_\nu(X,p)\mathcal{A}^\nu_{(0)}(X,p)\right]^{\frac{1}{2}}}. \tag{70}$$

Due to the condition (48), the four-vector $\mathcal{A}^\mu_{(0)}(X,p)$ is space-like, $\mathcal{A}^\mu_{(0)}(X,p)\mathcal{A}^{(0)}_\mu(X,p) < 0$, therefore $S_\mu(X,p)S^\mu(X,p) = -1$. The quantity $S^\mu(X,p)$ describes the mean spin orientation of the classical particles placed at the space-time point $X$ and having the four-momentum $p$. On the other hand, the magnitude of $\mathcal{A}^\mu_{(0)}(X,p)$, through Eq. (69), determines how many spin up and spin down particles are present in the system. Using Eq. (68), we find

$$p^\nu \partial_\nu S^\mu + M \partial_\nu M \partial^\nu_p S^\mu + \frac{\partial_\nu M}{M} p^\mu S^\nu - \varepsilon^{\mu\nu\alpha\beta} \partial_\nu \Phi p_\alpha S_\beta = 0, \tag{71}$$

and consequently

$$p^\nu \partial_\nu F_{\pm s}(X,p) + M \partial_\nu M \partial^\nu_p F_{\pm s}(X,p) = 0. \tag{72}$$

Hence, in a similar way to transport equations for QED [16], we find that the spin components decouple. Since both functions $F(X,p)$ and $\mathcal{A}^\mu_{(0)}(X,p)$ are on the mass-shell, we can write the analogous decomposition as in (63), namely

$$F_{\pm s}(X,p) = 2\pi \left\{ \frac{\delta(p^0 - E_p(X))}{2E_p(X)} f^+_{\pm s}(X,\mathbf{p}) + \frac{\delta(p^0 + E_p(X))}{2E_p(X)} \left[ f^-_{\mp s}(X,-\mathbf{p}) - 1 \right] \right\}. \tag{73}$$

Substituting Eq. (73) into Eq. (72), and integrating over $p^0$ gives

$$p^\mu \partial_\mu f^\pm_{\pm s}(X,\mathbf{p}) + M\partial_\mu M \partial_p^\mu f^\pm_{\pm s}(X,\mathbf{p}) = 0. \qquad (74)$$

By a direct comparison of (63) and (73), we find $f^\pm(X,\mathbf{p}) = \frac{1}{2}\left[f^\pm_{+s}(X,\mathbf{p}) + f^\pm_{-s}(X,\mathbf{p})\right]$. Therefore the functions $f^\pm(X,\mathbf{p})$ simply represent the spin-averaged densities. This has been given in agreement with the normalization of (65) and (66).

## 8. Consistency of the classical transport equations

The kinetic equations for the quark densities (62) and for the spin (68) follow from Eqs. (37) and (40) considered to the first order in $\hbar$. We have yet to examine (i) whether equations (38), (39) and (41) are consistent with these to the same order and (ii) whether the additional information for the angle $\Phi(X)$ appearing in (68) can be obtained. We note that this kind of checking has been done in the case of QED in [16].

First, in order to illustrate that there is no contradiction between (37) and (39) we substitute the quantities $\mathcal{A}_\mu$ and $\mathcal{S}_{\nu\mu}$ obtained from Eqs. (34) and (35), respectively into Eq. (39). In this way, we find the following equation

$$\hbar \left\{ \partial_\mu \mathcal{F}_{(0)} - \frac{p^\nu}{\sigma_{(0)}}\left[\partial_\mu\left(p_\nu \frac{\mathcal{F}_{(0)}}{\sigma_{(0)}}\right) - \partial_\nu\left(p_\mu \frac{\mathcal{F}_{(0)}}{\sigma_{(0)}}\right)\right] - \frac{\pi_{(0)}}{\sigma_{(0)}}\partial_\mu \mathcal{P}_{(0)} \right.$$
$$\left. + \frac{1}{\sigma_{(0)}}\left[\pi_{(0)}\partial_\nu \pi_{(0)}\partial_p^\nu \mathcal{V}^{(0)}_\mu + \sigma_{(0)}\partial_\nu \sigma_{(0)}\partial_p^\nu \mathcal{V}^{(0)}_\mu\right]\right\} = 0. \qquad (75)$$

Since we are interested in retaining terms to the first order in $\hbar$ we can use the relations (44), (45), (54), and the kinetic equation (62) to find that (75), and consequently (39), is fulfilled.

In an analogous fashion, using the kinetic equation for spin (68), the constraints (48) and (57), and the definition (61), we can show that Eq. (41) is fulfilled again to the first order in $\hbar$. To do this, we multiply both sides of (41) by $\varepsilon_{\alpha\beta\mu\nu}$. Using the formula for $\mathcal{V}^\nu$ from (33), we find that the multiplication of the left-hand-side of (41) gives the expression

$$-\frac{\hbar}{\sigma_{(0)}}\left\{\varepsilon_{\alpha\beta\mu\nu}p^\mu\left[\partial_\gamma \mathcal{S}^{\gamma\nu}_{(0)} - \partial_\gamma \pi_{(0)}\partial_p^\gamma \mathcal{A}^\nu_{(0)}\right] - \sigma_{(0)}\left(\partial_\alpha \mathcal{A}^{(0)}_\beta - \partial_\beta \mathcal{A}^{(0)}_\alpha\right)\right\}. \qquad (76)$$

On the other hand, the multiplication of the right-hand-side of (41) yields

$$\hbar \left[ \partial_\gamma \sigma_{(0)} \partial_p^\gamma \tilde{\mathcal{S}}_{\alpha\beta}^{(0)} + \partial_\gamma \pi_{(0)} \partial_p^\gamma \mathcal{S}_{\alpha\beta}^{(0)} \right]. \tag{77}$$

Using now the expressions for $\mathcal{S}_{\mu\nu}^{(0)}$ and $\tilde{\mathcal{S}}_{\mu\nu}^{(0)}$, after a rather lengthy calculation in which Eqs. (48), (57), (61) and (68) must be used, we find that the two expressions above are equal to each other. Consequently, Eq. (41) is fulfilled up to the first order in $\hbar$ as well.

In this way, we have shown that Eqs. (37), (39) - (41) are satisfied in the first order. To prove this fact, we have used Eqs. (31) - (35), treating them as if they were satisfied up to the first order. Is this procedure admissable? The answer to this question is yes. Only $\mathcal{F}$ and $\mathcal{A}^\mu$ are independent functions (with $\mathcal{A}^\mu$ restricted by the axial current conservation (36)). Eqs. (32), (33) and (35) are just the definitions of $\mathcal{P}$, $\mathcal{V}^\mu$ and $\mathcal{S}^{\mu\nu}$, and therefore are in our approach always satisfied. On the other hand Eq. (31) gives in the first order

$$(p^2 - M^2)\mathcal{F}_{(1)} = 2 \left[ \sigma_{(0)}\sigma_{(1)} + \pi_{(0)}\pi_{(1)} \right] \mathcal{F}_{(0)} + \partial_\nu \pi_{(0)} \mathcal{A}_{(0)}^\nu \tag{78}$$

and

$$(p^2 - M^2)\mathcal{P}_{(1)} = 2 \left[ \sigma_{(0)}\sigma_{(1)} + \pi_{(0)}\pi_{(1)} \right] \mathcal{P}_{(0)} + \partial_\nu \sigma_{(0)} \mathcal{A}_{(0)}^\nu, \tag{79}$$

whereas using Eq. (34) one finds

$$(p^2 - M^2)\mathcal{A}_{(1)}^\mu = 2 \left[ \sigma_{(0)}\sigma_{(1)} + \pi_{(0)}\pi_{(1)} \right] \mathcal{A}_{(0)}^\mu - M^2 F \partial^\mu \Phi. \tag{80}$$

One can notice that Eq. (80) as it stands is a chirally invariant expression. On the other hand Eqs. (78) and (79) form a system of chirally invariant equations in the same way as equations (26) and (27), or equations (29) and (30). One can also check, that using formula (32) as the definition of $\mathcal{P}$ and Eq. (78), we can derive Eq. (79). Therefore, only one of equations (78) and (79) is really independent. We observe that all the expressions (78) - (80) are just the generalized mass-shell constraints for $\mathcal{F}_{(1)}$, $\mathcal{P}_{(1)}$ and $\mathcal{A}_{(1)}^\mu$, and they do not influence the relations in the zeroth order.

The last equation that we are still required to check is Eq. (38). Substituting the quantity $\mathcal{P}$, calculated from (32) into (38), gives

$$\frac{M^2(X)}{2} \partial_\mu \Phi(X) \partial_p^\mu F(X,p) = p^\mu \mathcal{A}_\mu^{(1)}(X,p). \tag{81}$$

We see that Eq. (81) defines the parallel part of $\mathcal{A}^\mu$ in the first order. Multiplying (80) by $p^\mu$ and using Eqs. (81) and (54) we obtain zero. Hence Eqs. (81) and (80) are consistent.

After deriving all possible equations up to the first order in $\hbar$ and checking their consistency, we are still faced with two problems. Firstly, the system of our equations in the leading order is not closed, since we do not have an equation for $\Phi(X)$. Secondly, it is still not clear whether our kinetic equation for spin satisfies the requirement of the axial current conservation.

One *a priori* possible situation is that the condition of the axial current conservation eliminates the freedom connected with the choice of $\partial_\mu \Phi(X)$. However, by studying some simple situations, one can convince oneself that this is not the case. For example, assuming that $M = \text{const}$ and that initially the spin distribution is homogenous $\mathcal{A}^\mu_{(0)}(t=0, \mathbf{x}, p) = B^\mu(p)$ we find $p^0 \partial_0 \mathcal{A}^0_{(0)} = \varepsilon^{0ijk} \partial_i \Phi \, p_j \, B_k(p)$. In this situation, the axial current conservation at $t = 0$ requires that $\partial_i \Phi \int d^4p \, \varepsilon^{0ijk} (p_j/p^0) B_k(p) = 0$. We can see that this equation is not sufficient to determine the gradient of $\Phi$ and consequently the derivative $\partial_0 \mathcal{A}^0_{(0)}$ remains not well defined. Consequently, in order to obtain the closed system of equations in the classical limit we have to make some assumption on the form of $\partial_\mu \Phi(X)$. The simplest chirally invariant choice is

$$\partial_\mu \Phi(X) = 0. \tag{82}$$

In the case (82), our kinetic equation (68) allows us to determine the time evolution of the function $\mathcal{A}^\mu_{(0)}(X, p)$ from its knowledge at some initial time. One can notice, however, that this equation allows for solutions which do not satisfy the requirement of the axial current conservation. One of these solutions, for the case $M = \text{const}$, has the form $\mathcal{A}^\mu_{(0)}(X, p) = s^\mu(p) s_\nu(p) X^\nu \delta^{(4)}(p - \tilde{p})$, where $s^\mu(p)$ is a vector satisfying the conditions $s^\mu(p) p_\mu = 0$ and $s^\mu(p) s_\mu(p) = -1$, and $\tilde{p}$ is some given value of the four-momentum, with $\tilde{p}^\mu \tilde{p}_\mu = M^2$. On the other hand, there also exist solutions of Eq. (68) which satisfy the requirement of the axial current conservation. These are, e.g., the static homogeneous solutions (for the case $M = \text{const}$) or the solutions which are odd functions of momentum.

In consequence, we observe that Eq. (68) neither guarantees nor contradicts Eq. (10). Therefore the axial current conservation should be used as an external condition which selects the physical solutions of (68).

## 9. Explicit breaking of chiral symmetry

In the case $m_0 \neq 0$, the chiral invariance of the Lagrangian (1) is explicitly broken by the expression $-m_0 \bar{\psi} \psi$. In this situation, the additional term in the Dirac equation leads to a simple modification of our kinetic equations (24) - (28); we have to replace the mean field $\sigma$ by the sum $\sigma + m_0$. Of course, in practice, this change affects only the real parts of our equations, i.e., the formulae (31) - (35), since the imaginary parts (37) - (41) depend only on the gradients of $\sigma$. On the other hand, we observe that Eqs. (58) and (59) remain unchanged because they are just definitions of the mean fields.

The important fact in the non-symmetric case is that the relations between the functions $\mathcal{F}_{(0)}, \mathcal{P}_{(0)}$ and $\mathcal{V}_{(0)}^\mu$ look different. In particular, instead of Eqs. (52) and (53), we now find

$$\mathcal{P}_{(0)} = -\pi_{(0)} \frac{\mathcal{F}_{(0)}}{\sigma_{(0)} + m_0}, \quad \mathcal{V}_{(0)}^\mu = p^\mu \frac{\mathcal{F}_{(0)}}{\sigma_{(0)} + m_0}. \tag{83}$$

Using the first of the relations (83) in Eqs. (58) and (59), we obtain

$$\sigma_{(0)}(X) + 8G \int \frac{d^4p}{(2\pi)^4} \mathcal{F}_{(0)}(X, p) = 0, \tag{84}$$

and

$$\pi_{(0)}(X) \left[ \sigma_{(0)}(X) + m_0 + 8G \int \frac{d^4p}{(2\pi)^4} \mathcal{F}_{(0)}(X, p) \right] = 0. \tag{85}$$

A straightforward consequence of these two equations is the condition

$$\pi_{(0)}(X) = 0, \tag{86}$$

which implies also that $\mathcal{P}_{(0}(X, p) = 0$ and $\Phi(X) = 0$. Therefore, we observe that if the system is not chirally invariant, equations (58) and (59) allow for the determination of the angle $\Phi(X)$.

For $m_0 \neq 0$, the mass-shell constraints (54) and (57), as well as the kinetic equations (62) and (68) preserve their form, only in this case we should use the definitions

$$F(X, p) = \frac{\mathcal{F}_{(0)}}{\sigma_{(0)} + m_0}, \quad M(X) = \sigma_{(0)} + m_0. \tag{87}$$

The gap equation (65) has now the form

$$M = m_0 + 4GM \int \frac{d^3p}{(2\pi)^3} \frac{1}{\sqrt{M^2(X) + \mathbf{p}^2}} \left[ 1 - f^+(X, \mathbf{p}) - f^-(X, \mathbf{p}) \right]. \tag{88}$$

Let us now discuss the behavior of the axial current in the case when $m_0 \neq 0$. In this situation, we no longer have an axial current conservation law, and using Eq. (32), we can write

$$\hbar \left\{ \frac{1}{2} \partial_\mu \mathcal{A}^\mu_{(0)} - (\sigma_{(0)} + m_0) \mathcal{P}_{(1)} - \pi_{(1)} \mathcal{F}_{(0)} \right\} = 0. \tag{89}$$

The integration of (89) over momentum gives

$$\partial_\mu A^\mu_{(0)}(X) = -\frac{m_0}{G} \pi_{(1)}(X). \tag{90}$$

In contrast to the $m_0 = 0$, case we observe that such an integration does not lead to any constraint for $\mathcal{A}^\mu_{(0)}$ itself. Therefore, without any restrictions on $\partial_\mu \mathcal{A}^\mu_{(0)}$ we can treat Eq. (89) as a defining equation for $\mathcal{P}_{(1)}$. One can still check that Eq. (89) leads to the mass-shell constraint for $\mathcal{P}_{(1)}$ which agrees with (79). The latter formula, for the case $m \neq 0$, should be rewritten in the form $(p^2 - M^2)\mathcal{P}_{(1)} = \partial_\nu M \mathcal{A}^\nu_{(0)}$. We note that Eq. (90) describes the partial conservation of the axial current (PCAC) in our case.

## 10. Summary and Conclusions

In this paper, the chiral symmetry concepts have been explicitly included in the construction of the transport equations for quark matter. Our starting point was a chirally invariant Lagrangian and we derived the transport equations via a spinor decomposition of the Wigner function and a gradient expansion. Our calculation was restricted to the mean field approximation. We have taken into account the spin dynamics and discussed the possibility of having a non-zero pseudoscalar condensate. In this aspect, our results are a generalization of some earlier results obtained in Ref. [11].

The classical quark distribution functions satisfy the kinetic equations of the standard form with the effective chirally invariant mass $M^2(X) = \pi^2_{(0)}(X) + \sigma^2_{(0)}(X)$, where $\pi_{(0)}(X)$ and $\sigma_{(0)}(X)$ are leading terms in the classical expansion of the pseudoscalar and scalar condensates. Furthermore the angle $\Phi(X)$, defined through the relation $\pi_{(0)}(X)/\sigma_{(0)}(X) = \tan\Phi(X)$, must be a constant. However, its value remains undefined, which reflects the chiral symmetry of the problem.

The classical equation for the spin evolution, which has been derived here for the first time, is also invariant under chiral transformations. However, its solutions are constrained by the additional condition of the axial current conservation, that is not simple to incorporate.

The inclusion of the quark mass term into the Lagrangian explicitly breaks the chiral invariance. Consequently, we find that $\Phi(X)$ as well as $\pi_{(0)}(X)$ must be zero. Nevertheless, the general form of the kinetic equations does not change and we can still use them with the substitution $M(X) = \sigma_{(0)}(X) + m_0$, where $m_0$ is the current quark mass. Moreover, in this case the requirement of the axial current conservation reduces to a form familiar from PCAC.

*Acknowledgements:* We thank M. Volkov, F. Witte and P. Zhuang for clarifying discussions.